\documentclass[10pt]{article}
\usepackage{multicol}
\usepackage{amssymb}
\usepackage{amsfonts,amssymb,amsmath,amsthm}
\usepackage{epsfig}
\usepackage{slashed}
\usepackage{graphicx}
\usepackage{float}
\usepackage{makecell}
\usepackage{mathrsfs}
\usepackage{bbm}
\usepackage{cite}
\usepackage{times, mathptmx}
\usepackage{enumerate}
\usepackage{longtable}
\usepackage{booktabs}
\usepackage[utf8]{inputenc}
\usepackage[colorlinks,linkcolor=red,urlcolor=cyan,anchorcolor=blue,citecolor=blue]{hyperref}
\usepackage{titlesec}
\titleformat{\section}
{\raggedright\large\bfseries}
{\thesection .}
{0pt}
{}

\renewcommand\thesection{\arabic{section}}

\begin{document}

\begin{center}
{\large \bf Scalar quasinormal modes of black holes in Einstein-Yang-Mills gravity}
\vspace{4mm}

{{ Yang Guo${}^{}$\footnote{\em E-mail: guoy@mail.nankai.edu.cn}
and Yan-Gang Miao}${}^{}$\footnote{\em Corresponding author.}${}^{,}$\footnote{\em E-mail: miaoyg@nankai.edu.cn}

\vspace{1mm}
${}^{}${\normalsize \em School of Physics, Nankai University, Tianjin 300071, China}
}
\end{center}

\noindent
Quasinormal modes of a scalar field perturbation are investigated in the background spacetime of Einstein-Yang-Mills black holes. 
%There are different effects on Hawking temperature and quasinormal modes from the logarithmic term . 
The logarithmic term in the metric function of the five-dimensional Einstein-Yang-Mills black hole eliminates the divergence of the Hawking temperature but maintains the charge-dependent behavior of the quasinormal modes. The real and imaginary parts of quasinormal mode frequencies have the same charge-dependent behavior in different dimensions. Similar to the case of high dimensional Schwarzschild black holes, a compact approximate relation, $\omega_{\rm R}\sim D/r_{+}$, also exists between the real part of the quasinormal mode frequencies $\omega_{\rm R}$ and the number of dimensions $D$ in the higher than five dimensional Einstein-Yang-Mills black holes.

\vspace{5mm}

\renewcommand{\thefootnote}{\arabic{footnote}}
\setcounter{footnote}{0}
\setcounter{page}{2}
\pagenumbering{arabic}

\section{ Introduction}
Quasinormal modes (QNMs) of black holes, which characterize the dynamics of perturbation fields, are usually used to describe a dissipative system that oscillates and decays at a complex frequency, $\omega=\omega_{\rm R}-i\omega_{\rm I}$, where the real part represents the frequency of oscillation while the imaginary part gives the damping timescale. For a stable black hole, the imaginary part is negative, i.e. $\omega_{\rm I}$ is positive. The fundamental mode $(n=0)$ is the least damped and long lived mode in a ringdown signal, which is most likely to be detected in a binary compact star merger.

The investigations of QNMs have been made in a wide range, see, for instance, some reviews~\cite{HPN01,KDKS04,EBCS02,RAKAZ03}. In the Einstein theory and Einstein-Maxwell theory, the QNMs of a scalar field perturbation have been studied extensively, such as for the Reissner-Nordstr\"om-anti-de Sitter (RNAdS) black hole~\cite{EBK07}, Kerr black hole~\cite{RAKZ06}, and Reissner-Nordstr\"om (RN) black hole~\cite{MRG05}, respectively. In particular, it has been found~\cite{RAK08} that the real part of QNM frequencies is proportional to the ratio of the spacetime dimension to the horizon radius, $D/r_{+}$, in high dimensional Schwarzschild black holes. 
%The thermalization timescale in the strong coupled conformal field theory (CFT) was determined~\cite{GTHH09} by the QNMs of a minimally coupled scalar field. 
Furthermore, the timescale has been computed~\cite{GTHH09,IS} numerically via the QNMs of a scalar field perturbation in different spacetime dimensions because the timescale from a perturbed state to a thermal equilibrium can be characterized by the inverse of imaginary parts of frequencies.. 

Compared with the Einstein-Maxwell theory, the Einstein-Yang-Mills (EYM) theory is more challenging in finding exact solutions due to its inherent complications. Until now, the numerical solutions have been available for the $SU(N)$ EYM theory, where the solutions of the $SU(2)$ EYM theory are asymptotically flat and uncharged~\cite{RBM,PB} while that of the $SU(5)$ EYM theory, in contrast to the case in the $SU(2)$ EYM theory, are charged~\cite{BKKS10}. It has been shown~\cite{NSZ,ZHZS,MSVBLS} that these solutions are unstable in terms of numerical analyses under linear perturbations. Nonetheless, the black hole solutions in the four dimensional $SU(N)$ EYM theory with a negative cosmological constant have been proved~\cite{JEBW} to be stable. Thus, the EYM black holes in the higher than four dimensions are quite different from that in the four dimensions. 

The analytical solutions have been obtained~\cite{SHMH11,SHMH12,YBCHT,NOM,ACT,YBRT} when the symmetric group is reduced to $SO(D-1)$, where the order of the gauge group is less than the number of dimensions by one. However, the stability of these solutions under linear perturbations remains to be verified. The significant difference between the $SO(D-1)$ EYM black hole and the Einstein-Maxwell RN black hole is the appearance of an additional logarithmic term in the metric function of the five-dimensional EYM black hole, which leads to~\cite{SHMH11} a mass-independent radius of horizon. What we are interested in is whether there are new physical effects caused by the logarithmic term, in particular, how it affects QNMs and what behaviors the QNMs exhibit in different dimensions. We are going to investigate the behavior of QNMs determined by the parameters of black holes and in particular by a gauge charge. The issue is dealt with by a massless scalar field around EYM black holes in various dimensions. It is usually interesting if a rule that is valid in high dimensional Schwarzschild and Reissner-Nordstr\"om black holes can be generalized to EYM black holes or not. We focus on the research of QNMs and find that the relationship between QNMs and the number of dimensions will be present in the EYM black holes with the $SO(D-1)$ gauge symmetry. In addition, we are able to gain some insight into the stability of black holes in the Einstein gravity coupled to Yang-Mills gauge fields. Specifically, we want to determine the ranges of Yang-Mills charges and spacetime dimensions for stable EYM black holes. Therefore, we shall study the EYM black holes with the $SO(D-1)$ gauge symmetry in the higher than four dimensions due to their specifics mentioned above. 

%We investigate these issues by dealing with a scalar field around EYM black holes and applying the analytical solutions of EYM black holes with the $SO(D-1)$ gauge group. Our aim is to gain some insight into the stability of black holes in Einstein gravity coupled to Yang-Mills gauge fields. Specifically, we want to determine the regions of Yang-Mills charges and dimensions for stable EYM black holes. We shall focus on the EYM black holes with the $SO(D-1)$ gauge group in the higher than four dimensions due to their specifics. 

The outline of this paper is as follows. In Sec.~\ref{sec:EYM-BH} we at first review shortly the EYM black holes, and then investigate the Hawking temperature of the EYM black holes in $D=5$ and $D=6$ as an example of $D>5$. Sec.~\ref{sec:QNM} contains the derivation of the perturbation equations of a massless scalar field for a static spherically symmetric EYM black hole, and is divided into three subsections, where the first two subsections are devoted to the cases of $D=5$ and $D > 5$, respectively, and the last one to the effects of a varying charge and the numerical convergence of the WKB method. In Sec.~\ref{sec:d=5}, we compute the fundamental QNMs under a massless scalar field perturbation 
%analyze their differences from that of the RN black hole, 
and study the dependence of QNMs on gauge charges and angular quantum numbers for the case of $D=5$. Next, we 
%make the comparative investigations for our results and 
analyze mainly the dependence of QNMs on the number of dimensions for the case of $D>5$ in Sec.~\ref{sec:d>5}. 
Since the relation between $\omega_{\rm R}$ and $D/r_{+}$ is revealed only for a fixed charge in the above subsection, we investigate it for a varying charge in conjunction with the numerical convergence of the WKB method in Sec.~\ref{sec:limit}.
Finally, we present our conclusions in Sec.~\ref{sec:conclusions}.

\section{ Einstein-Yang-Mills black holes }\label{sec:EYM-BH}
The EYM gravity we consider here is described~\cite{SHMH11,SHMHA13} by the action,
\begin{eqnarray}\label{action}
S=\frac{1}{2}\int\limits_\mathcal{M}  {\rm d}^D{x} \sqrt{-g}\left[ R-\sum_{a=1}^{(D-1)(D-2)/2} F_{\mu\nu}^{a} F^{a\mu\nu}\right],
\end{eqnarray}
where $R$ is the Ricci scalar, $F_{\mu\nu}^{a}=\partial_{\mu}A_{\nu}^{a}-\partial_{\nu}A_{\mu}^{a}+\frac{1}{2\sigma}C_{bc}^{a}A_{\mu}^{b}A_{\nu}^{c}$ are the field strengths of $SO(D-1)$ Yang-Mills fields with the structure constants $C_{bc}^{a}$ and coupling constant $\sigma$,  and $A_{\mu}^{a}$ are the gauge potentials, the Latin indices, $a, b, c, \dots=1, 2, \dots, (D-1)(D-2)/2$, represent the internal space of the gauge group, and the Greek indices, $\mu, \nu, \alpha, \beta, \dots=0, 1, \dots, D-1$, describe the $D$-dimensional spacetime. The EYM field equations can be obtained\footnote{The Yang-Mills field equations can be derived by the variation of the action with respect to the gauge potentials $A_{\mu}^{a}$. They are not written down due to their irreverence to the discussions below.} by the variation of the action with respect to the metric $g_{\mu\nu}$,
\begin{eqnarray}\label{EYM-equation}
G_{\mu\nu}=T_{\mu\nu},
\end{eqnarray}
where the energy-momentum tensor can be expressed by
\begin{eqnarray}
T_{\mu\nu}=\sum_{a=1}^{(D-1)(D-2)/2}\left[2F_{\mu}^{a\lambda}F_{\nu\lambda}^{a}-\frac{1}{2}F_{\alpha\beta}^{a}F^{a\alpha\beta}g_{\mu\nu} \right].
\end{eqnarray}

The following intuitive result can be obtained~\cite{SHMH11,SHMH14} in terms of the $D$-dimensional Wu-Yang ansatz, $A^{a}\equiv A^{a}_{\mu}{\rm d}x^{\mu}=\frac{Q}{r^2}(x_i{\rm d}x_j-x_j{\rm d}x_i)$, $2 \le i \le D-1$, $1\le j\le i-1$, 
% $2\leq i\leq d-1$, $1\leq j\leq i-1$, $1\leq (a)\leq (d-1)(d-2)/2$ 
and the only non-zero gauge charge $Q$,
\begin{eqnarray}
\sum_{a=1}^{(D-1)(D-2)/2}[F_{\alpha\beta}^{a}F^{a\alpha\beta}]=\frac{(D-3)(D-2)Q^2}{r^4},
\end{eqnarray}
which yields the non-zero components of the energy-momentum tensor.
% Compared with familiar Einstein-Maxwell theory, the intrinsic complication of EYM theory makes it necessary to specify to spherical symmetry.
As a result, the $D$-dimensional line element takes a general form,
\begin{eqnarray}
{\rm d}s^2=-f(r){\rm d}t^2+\frac{{\rm d}r^2}{f(r)}+r^2{\rm d}\Omega_{D-2}^2,
\end{eqnarray}
where ${\rm d}\Omega_{D-2}^2 $ is the line element of the unit sphere $S^{D-2}$. Correspondingly, Eq.~(\ref{EYM-equation}) reduces to two radial equations in five dimensions and higher than five dimensions, respectively, based on which the metric functions can be expressed as
\begin{alignat}{2}
&	f(r)=1-\frac{m}{r^2}-\frac{2Q^2}{r^2}\ln(r) &\qquad{(D=5)}, \label{EYM-metric} \\
&f(r)=1-\frac{m}{r^{D-3}}-\frac{(D-3)Q^2}{(D-5)r^2} &\qquad{(D>5)},\label{EYM-metric2}
\end{alignat}
where $m$ denotes the mass of black holes. Incidentally, the two metric functions can alternatively be derived~\cite{SHMH14} in $f(R)$ gravity when $f(R)=R$.

\begin{figure}[H]
	\centering
	\includegraphics[width=0.45\linewidth]{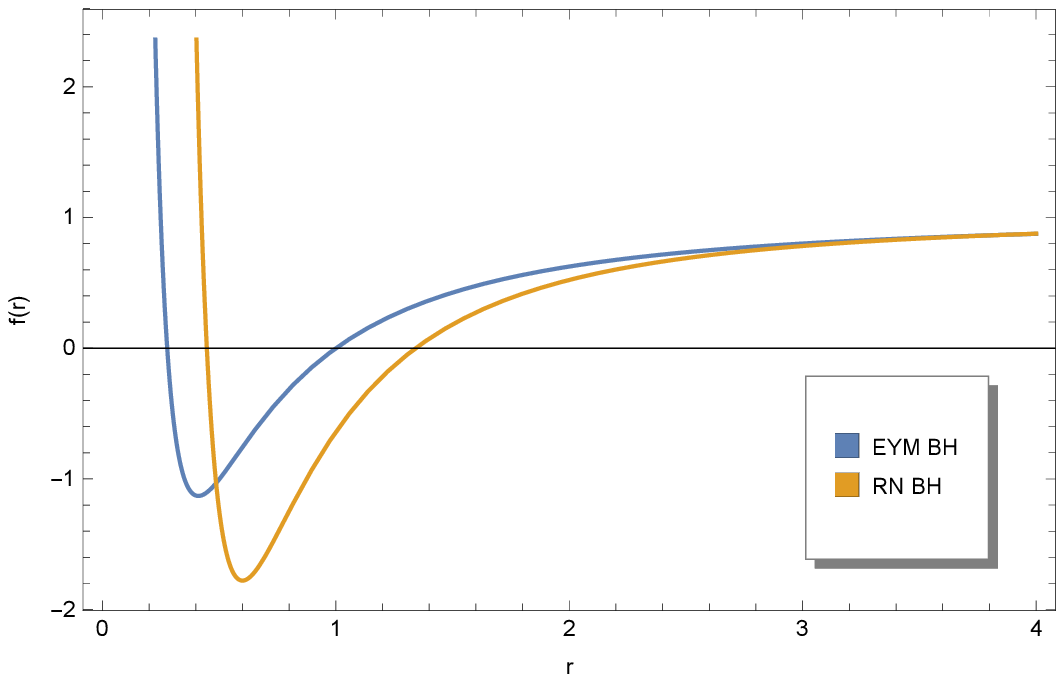}  \quad \includegraphics[width=0.45\linewidth]{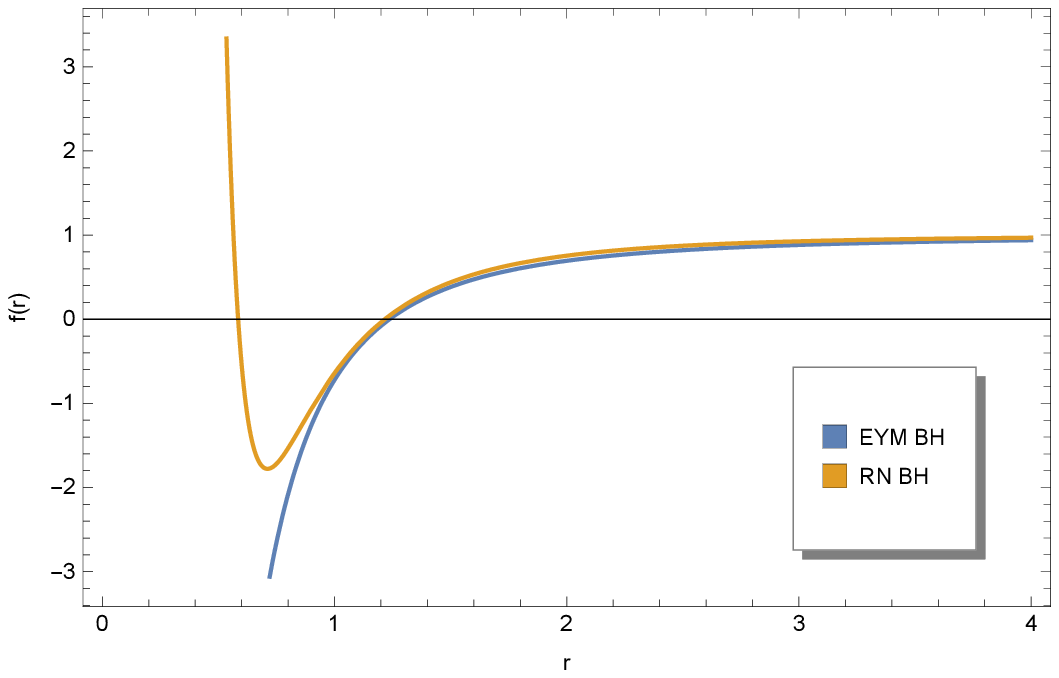}
	\caption{The figure shows the metric function $f(r)$ for the Einstein-Yang-Mills black hole and Reissner-Nordstr\"om black hole in $D=5$ (left) and $D=6$ (right) with the fixed black hole mass $m=1$ and gauge charge $Q=0.6$.
	} \label{fig:f(r)}
\end{figure}

The two metric functions of the Einstein-Yang-Mills black holes are plotted in Fig.~\ref{fig:f(r)}, where the metric functions of the Reissner-Nordstr\"om black holes are attached for comparison. It is obvious from the figure that the behavior of the metric function of the EYM black hole with respect to the radial coordinate is similar to that of the metric function of the RN black hole in $D=5$, and that the EYM black hole has only one event horizon but the RN black hole has one event horizon and one Cauchy horizon in $D=6$. As the number of dimensions increases, the positions of the two horizons in the RN black holes approach to each other and coincide at $r=1$, while the position of the event horizon in the EYM black holes approaches one from outside. We note that the metric function of the four-dimensional Bardeen black hole converges~\cite{SFC15} to one in the limits of $r \to 0$ and $r\to\infty$. However, the metric functions of the EYM black holes diverge at $r\to0$ for both cases of $D=5$ and $D>5$, and only at infinity do the two functions converge to one. That is, we want to emphasize that the singularity remains for the EYM black holes in contrast to the non-singular nature of the Bardeen black hole.

For the given event horizon $r_+$, the Hawking temperature of the Einstein-Yang-Mills black holes and Reissner-Nordstr\"om black holes  can be obtained in terms of the formula, $T=\left.-\frac{1}{4\pi}\frac{{\rm d}g_{tt}}{{\rm d}r}\right|_{r=r_+}$, as follows:
\begin{alignat}{2}\label{T-5d}
&T_{\rm EYM}=\frac{1}{2\pi}\left[ \frac{m}{r_+^3}-\frac{Q^2}{r_+^3}+\frac{2Q^2\ln r_+}{r_+^3}\right]  &\qquad{(D=5)}, \\
&T_{\rm EYM}=\frac{1}{4\pi}\left[ \frac{3m}{r_+^4}+\frac{6Q^2}{r_+^3}\right] &\qquad{(D=6)},\\
&T_{\rm RN}=\frac{1}{\pi}\left[ \frac{m}{r_+^3}-\frac{Q^2}{r_+^5}\right] &\qquad{(D=5)},\\
&T_{\rm RN}=\frac{1}{2\pi}\left[ \frac{3m}{r_+^4}-\frac{3Q^2}{r_+^7}\right] &\qquad{(D=6)}.	
\end{alignat}

\begin{figure}[H]
	\centering
	\includegraphics[width=0.45\linewidth]{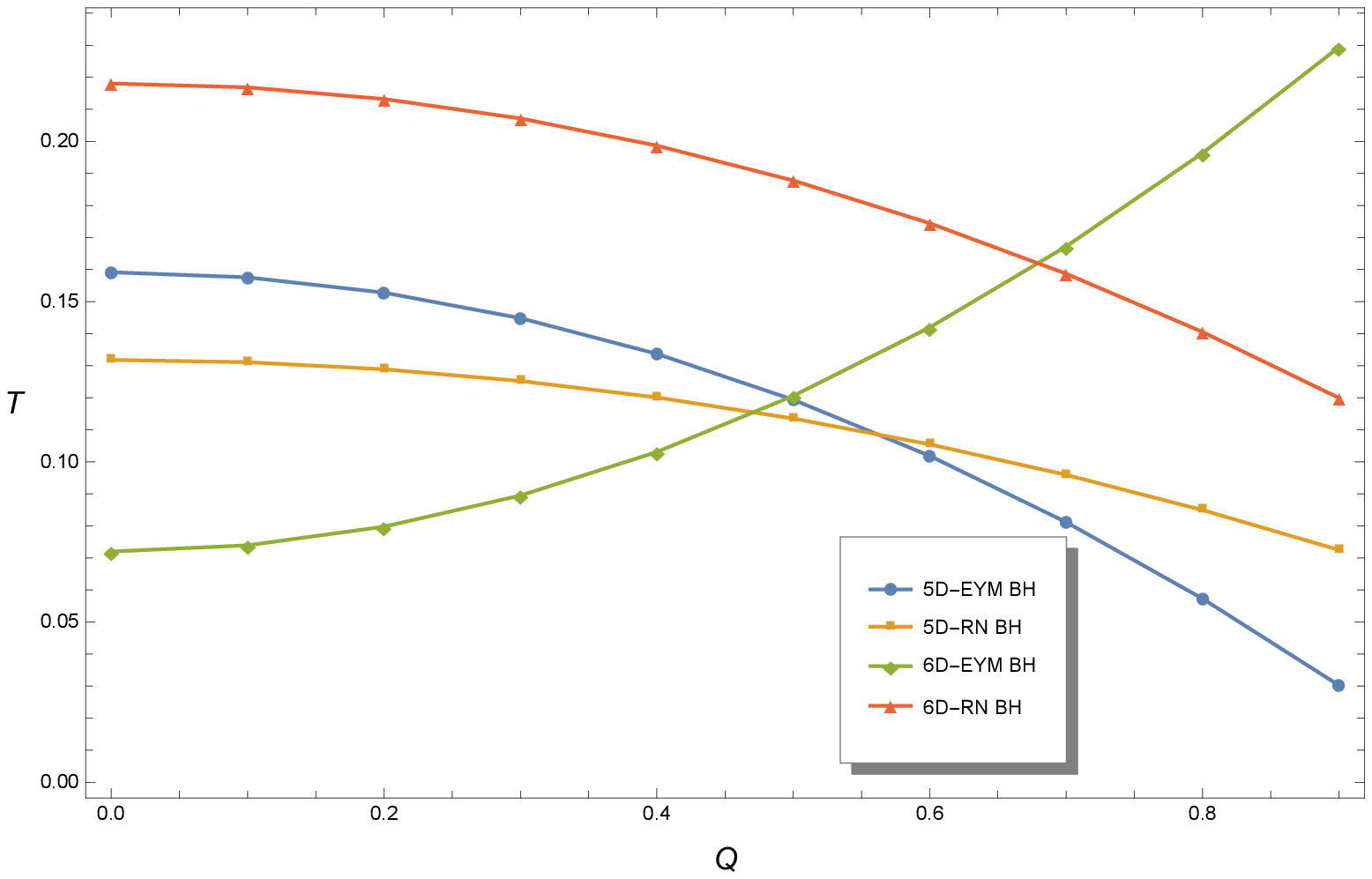}\label{fig:T-Q}  \quad \includegraphics[width=0.45\linewidth]{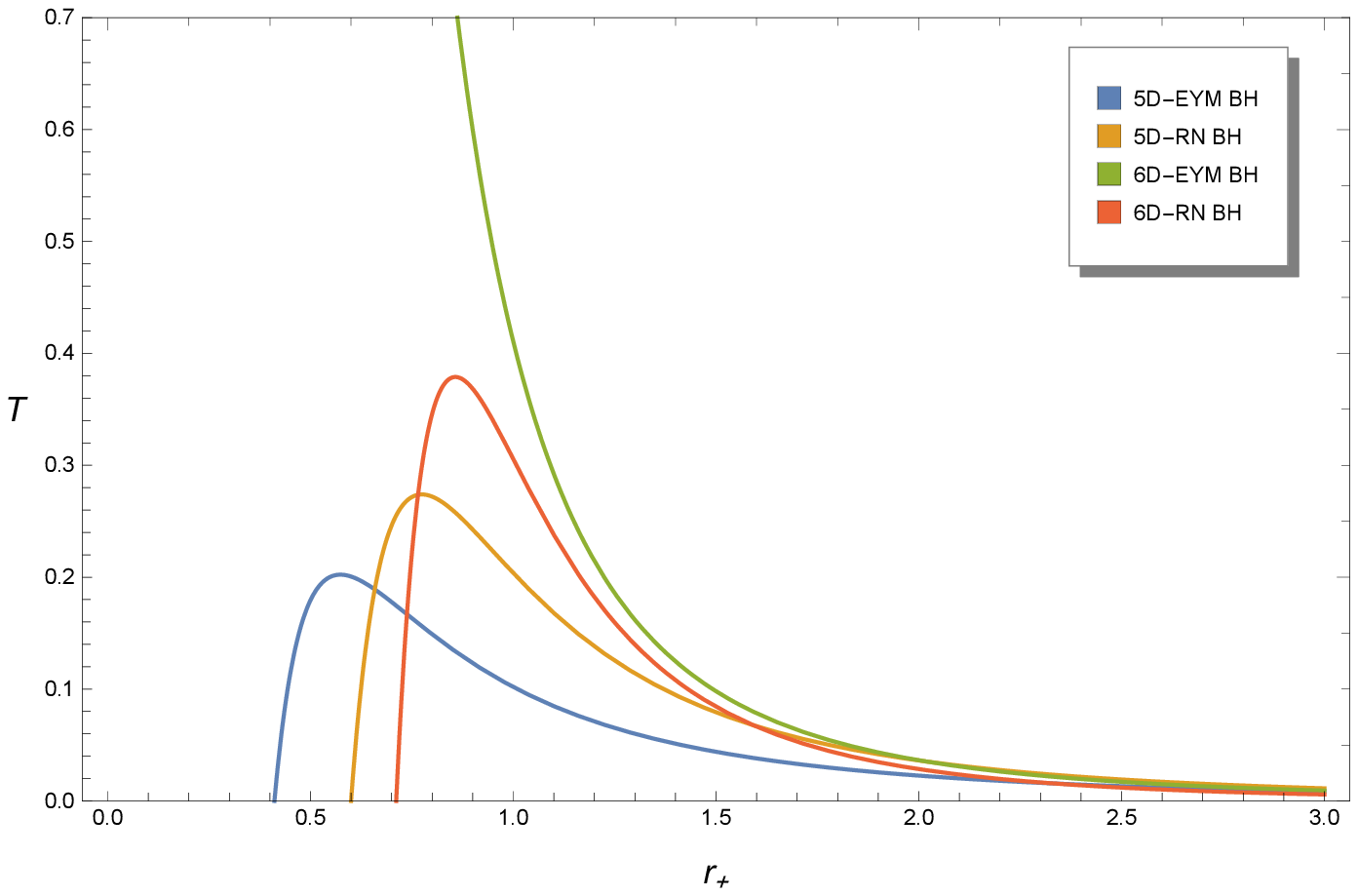}\label{fig:T-r}
	\caption{(left) The Hawking temperature of EYM black holes and RN black holes as a function of the gauge charge $Q$ in $D=5$ and $D=6$, where the black hole mass is fixed at $m=1$. (right) The profile of the Hawking temperature on the $r_+-T$ plane, where $m=1$ and $Q=0.6$ in $D=5$ and $D=6$.
	} \label{fig:T}
\end{figure}

The Hawking temperature is plotted in Fig.~\ref{fig:T} with respect to the gauge charge $Q$ and the event horizon $r_+$, respectively.
For the case of $D=5$, the temperatures of both the EYM black hole and RN black hole decrease when the gauge charge $Q$ increases, see the left diagram. Their difference is stated as follows: The temperature of the EYM black hole is higher than that of the RN black hole when $Q$ is smaller than $0.6$, but it is lower than the temperature of the RN black hole when $Q$ is larger than $0.6$. Moreover, it is also clear for the case of $D=5$, see the right diagram, that the extreme temperatures exist during the evaporation of the EYM black hole and the RN black hole. Nevertheless, the most significant difference between the EYM black hole and the RN black hole is that the temperature of the EYM black hole ascends while the temperature of the RN black hole descends when $Q$ increases in the case of $D=6$, see the left diagram. In addition, also for the case of $D=6$, there is one specific feature worth noting: The evaporation of the EYM black hole will cause the divergence of the Hawking temperature, see the green curve of the right diagram, because no logarithmic term appears in the metric function Eq.~({\color{red}9}). In brief, we explicitly show the relations between the Hawking temperature and the charge in the left diagram of Fig.~\ref{fig:T} and the relations between the Hawking temperature and the horizon in the right diagram of Fig.~\ref{fig:T}. It is interesting to note that it is actually the logarithmic term that eliminates the temperature's divergence and makes the two black holes behave similarly in the case of $D=5$.

\section{ Quasinormal modes of  Einstein-Yang-Mills black holes under a massless scalar field perturbation}\label{sec:QNM}

The massless scalar field $\Phi$ propagating in a curved spacetime is governed by the Klein-Gordon wave equation,
\begin{eqnarray}
g^{\mu\nu}\nabla_\mu\nabla_\nu\Phi=0. \label{KGEq.}
\end{eqnarray}
For a stationary and spherically symmetric solution, we substitute the following decomposition of variables,
\begin{eqnarray}
\Phi(t,r,\theta,\varphi)=\sum_{l,m}e^{-i\omega t}\frac{\psi_{l}(r)}{r^{(D-2)/2}}Y_{lm}(\theta, \varphi),
\end{eqnarray}
%\begin{eqnarray}
%\phi(t,r,\Theta)=\sum_{lm}e^{-i\omega t}r^{(2-D)/2}\psi(r)Y_{lm}(\Theta),
%\end{eqnarray}
%into Eq.~(\ref{KGEq.}), where $Y_{lm}(\Theta)$ stands for the sperical harmonics on $S^{D-2}$ with the eigenvalue $-l(l+D-3)$, and then derive the radial equation for $\psi(r)$,
into Eq.~(\ref{KGEq.}), 
%in which the radial and angular dependence of the massless scalar field are reflected by $\psi(r)$ and 
where $Y_{lm}(\theta, \varphi)$ stands for the spherical harmonics of $D-2$ degrees, $0\le \theta=(\theta_1, \theta_2, \dots, \theta_{D-3})\le \pi$ and $0 \le \varphi\le 2\pi$, and then derive the radial equation for $\psi_l(r)$,
\begin{eqnarray}
f^2\psi''_l+ff'\psi'_l+\omega^2-\left[\frac{l(l+D-3)}{r^2}+\frac{(D-2)(D-4)}{4r^2}f+\frac{D-2}{2}f' \right]f \psi_l=0, \label{radialEq.}
\end{eqnarray}
where $f'\equiv {\rm d}f(r)/{\rm d}r$, $\psi'_l\equiv {\rm d}\psi_l(r)/{\rm d}r$, and $\psi''_l\equiv {\rm d}^2\psi_l(r)/{\rm d}r^2$. After defining the ``tortoise" coordinate by the relation,  ${\rm d}r_*={\rm d}r/f(r)$, we reduce the radial equation Eq.~(\ref{radialEq.}) into the standard form,
\begin{eqnarray}
\left[ \partial_{r_*}^2+\omega^2-V(r)\right] \psi_l(r)=0,\label{ODE}
\end{eqnarray}
%with the effective potentials is give by 
%\begin{eqnarray}
%	V(r)=f(r)\left[\frac{l(l+d-3)}{r^2}+\frac{(d-2)(d-4)}{4r^2}f(r)+\frac{d-2}{2}f'(r)  \right].
%\end{eqnarray}
where the effective potentials of the Einstein-Yang-Mills black holes for the cases of $D=5$ and $D>5$ can be calculated from the metric functions Eqs.~(\ref{EYM-metric}) and (\ref{EYM-metric2}),  
%\begin{subequations}%\label{EYM-potential}
	\begin{eqnarray}
	V(r)=\left(1-\frac{m}{r^2}-\frac{2Q^2\ln(r)}{r^2}\right)\left[ \frac{4l(l+2)+3}{4r^2}+ \frac{9m-12Q^2}{4r^4}+\frac{3Q^2\ln(r)}{2r^4} \right]  \qquad{(D=5)},\label{EYM-potential} \\
	\begin{split}
	V(r)=\left(1-\frac{m}{r^{D-3}}-\frac{(D-3)Q^2}{(D-5)r^2}\right)\bigg[\frac{4l(l+D-3)+(D-2)(D-4)}{4r^2}+\frac{m(D-2)^2}{4r^{D-1}} \hspace{0.6cm}\qquad\\
	-\frac{(D-2)(D-3)(D-8)Q^2}{4(D-5)r^4}   \bigg] \qquad{(D>5)}.\label{EYM-potential2}
	\end{split}
	\end{eqnarray}
%\end{subequations}

Hence, the evolution of the linear perturbation field $\Phi$ in the black hole spacetime turns to the differential equation Eq.~(\ref{ODE}), based on which the quasinormal modes can be determined. Moreover, we have to impose the following boundary conditions in order to obtain the solution of Eq.~(\ref{ODE}),
\begin{eqnarray}
\psi_l\sim e^{-i\omega(t\mp r_{*})},\qquad r_*\to\pm\infty.
\end{eqnarray}

For an asymptotically flat spacetime, such as the EYM black hole spacetimes, we see that only the outgoing modes can be present at $r_*\to +\infty$. In order to find the QNMs characterizing the dynamics of a massless scalar field, we apply the improved WKB approximation which is also called the higher order WKB-Pad\'e approach suggested first in Ref.~\cite{JMMO16} and developed later in Refs.~\cite{RAKZZ,JMT}. The Pad\'e approximation greatly improves the accuracy of computations based on the standard WKB method\cite{WKB} and we adopt the 13th order WKB-Pad\'e approach in our work. Note that the fundamental mode ($n=0$) is very long lived\footnote{For instance, the dominant mode of the ringdown waveform of astrophysical black holes is just the fundamental mode~\cite{EBCS02}.} when compared to the other modes with the overtone number $n\ge 1$, so we mainly discuss it in this paper. 
% ($n=0$, $l=2$) with the complex frequency 0.3737-0.0890i in geometric units. In this paper, the fundamental mode ($Q=0.6$) with complex frequency 1.297374-0.252624i (see Table~\ref{tab:table1}) gives a damping timescale, which is much bigger than the case of mode ($l=2$, $n=1$) with the complex frequency 1.242038-0.772118i that we didn't list in Table. 
Next, we study the fundamental modes for the EYM black holes with the $SO(D-1)$ gauge symmetry in $D=5$ and $D>5$, respectively, and extend the relation between $\omega_{\rm R}$ and $D/r_{+}$ from the special case of a fixed charge to the general case of a varying charge.

\subsection{The case of \boldmath$D=5$}\label{sec:d=5}

In accordance with Eqs.~(\ref{ODE}) and (\ref{EYM-potential}) and the numerical method~\cite{JMMO16}, we work out the fundamental QNM frequencies of a massless scalar field perturbation in the ranges of the gauge charges $Q=0.0, 0.1, \dots, 0.6$ and of the angular quantum numbers $l=0, 1, 2, 3, 4$, where the black hole mass is set to be unit. The results showing the dependence of QNMs on gauge charges and angular quantum numbers are listed in Table~\ref{tab:table1}.

\begin{table}[H]
	\caption{\label{tab:table1}The fundamental QNM frequencies $(n=0)$ of a massless scalar field perturbation in the five-dimensional EYM black hole for different values of $Q$ and $l$.}
	\vskip 2mm
	\resizebox{\textwidth}{!}{
	\begin{tabular}{ c ccccc}
		\hline
		\hline
		\textrm{$Q$}&	\textrm{$l=0$}&$l=1$&$l=2$&$l=3$&$l=4$\\
		\hline
		0.0 & 0.533677-0.383416$i$& 1.016016-0.362329$i$& 1.510567-0.357537$i$&2.007886-0.355802$i$&2.506291-0.354993$i$\\
		0.1 & 0.530591-0.381297$i$& 1.011766-0.359184$i$& 1.504823-0.354418$i$&2.000538-0.352738$i$&2.497287-0.351955$i$\\
		0.2 & 0.520853-0.370485$i$& 0.998915-0.349397$i$& 1.487527-0.345118$i$&1.978402-0.343594$i$&2.470160-0.342887$i$\\
		0.3 & 0.504477-0.351362$i$& 0.977465-0.333452$i$& 1.458492-0.329802$i$&1.941209-0.328530$i$&2.424561-0.327949$i$\\
		0.4 & 0.482955-0.330156$i$& 0.947146-0.311584$i$& 1.417409-0.308803$i$&1.888518-0.307872$i$&2.359919-0.307457$i$\\
		0.5 & 0.452094-0.295107$i$& 0.907866-0.284435$i$& 1.363856-0.282719$i$&1.819739-0.282202$i$&2.275482-0.281986$i$\\
		0.6 & 0.416501-0.258893$i$& 0.859046-0.253097$i$& 1.297374-0.252624$i$&1.734254-0.252572$i$&2.170474-0.252579$i$\\
		\hline
		\hline
	\end{tabular}}
\end{table}

\begin{figure}[htbp]
	\centering
	\includegraphics[width=0.45\linewidth]{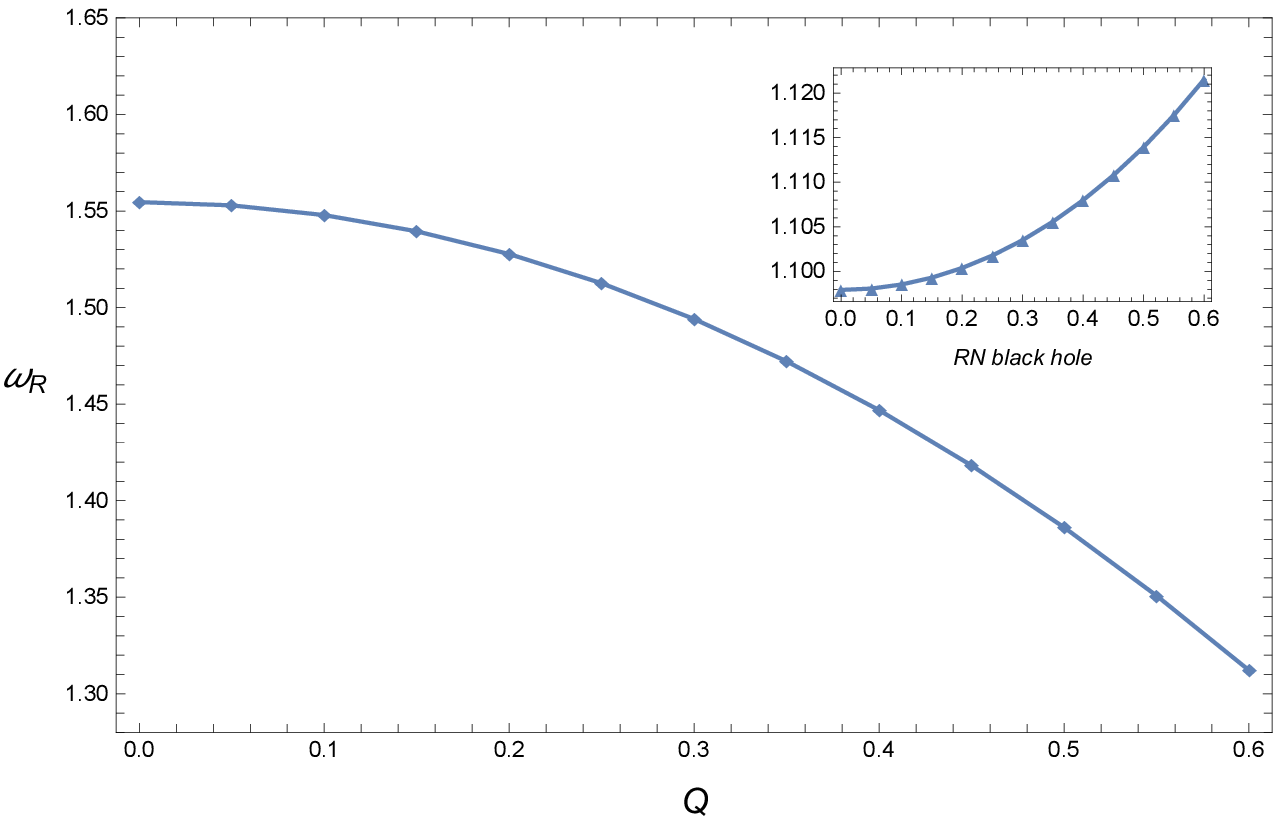}\label{fig:5dEYM(RN)Re}  \quad \includegraphics[width=0.45\linewidth]{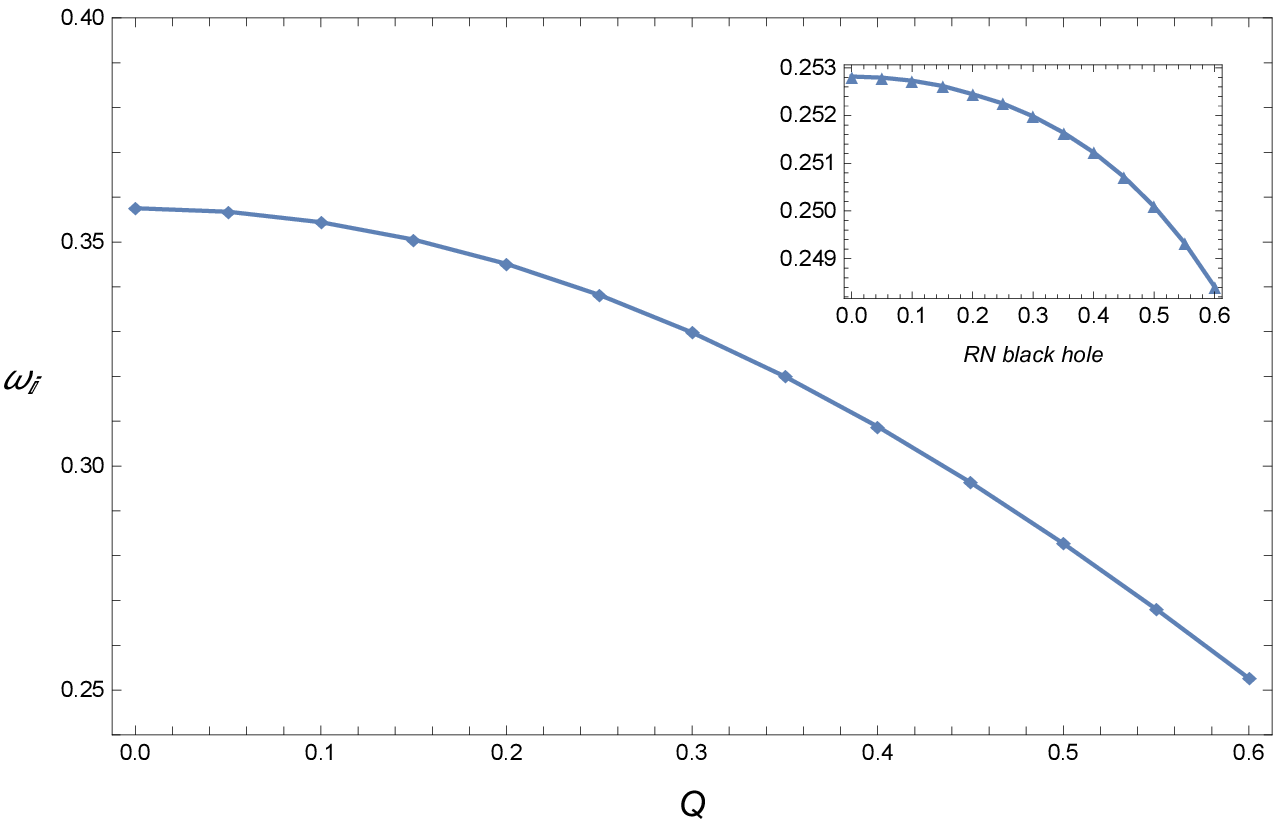}\label{fig:6dEYM(RN)Re}
	\caption{The behaviors of real parts (left) and imaginary parts (right) of QNM frequencies in the fundamental mode $(n=0, l=2)$ with respect to the charge $Q$ for the EYM black hole and RN black hole in five dimensions.
	} \label{fig:5DRe}
\end{figure}

In order to analyze in detail the behaviors of QNMs with respect to the gauge charge $Q$ in the EYM black hole, we compute more data for the fundamental mode ($n=0$) with the angular quantum number $l=2$ besides those in Table~\ref{tab:table1} and plot the frequencies with respect to the charge in Fig.~\ref{fig:5DRe}, where the case of the RN black hole is attached for comparison. 

%Fig.~\ref{fig:5DRe} gives us a comparison of the frequencies computed from the above approach between the EYM black hole and RN black hole.
From the left diagram of Fig.~\ref{fig:5DRe} we can see that the real part of frequencies decreases in the EYM black hole when the Yang-Mills charge increases, which is in contrast to the behavior of the real part of frequencies with respect to the Maxwell charge in the RN black hole.    
We note that the logarithmic term in the EYM black hole leads to the similar Hawking temperature profile to  that of the RN black hole in five dimensions in section~\ref{sec:EYM-BH}. However, the fact that the logarithmic term does not make the EYM black hole behave as the RN black hole does in five dimensions suggests that its effect on QNMs seems to be greatly weakened. 
In addition, from the right diagram of Fig.~\ref{fig:5DRe} we can see that the imaginary parts of QNM frequencies behave similarly, i.e. they decrease as $Q$ increases in both the EYM and RN black holes.\footnote{It is interesting to note~\cite{SFC15,EWL17,RAK18} that the imaginary part of frequencies in the four-dimensional RN black hole increases at first to its maximum value and then decreases with  the increasing of the Maxwell charge. However, the imaginary parts of frequencies in higher than four dimensional EYM and RN black holes have no maximum values.}  

\begin{figure}[H]
	\centering
	\includegraphics[width=0.45\linewidth]{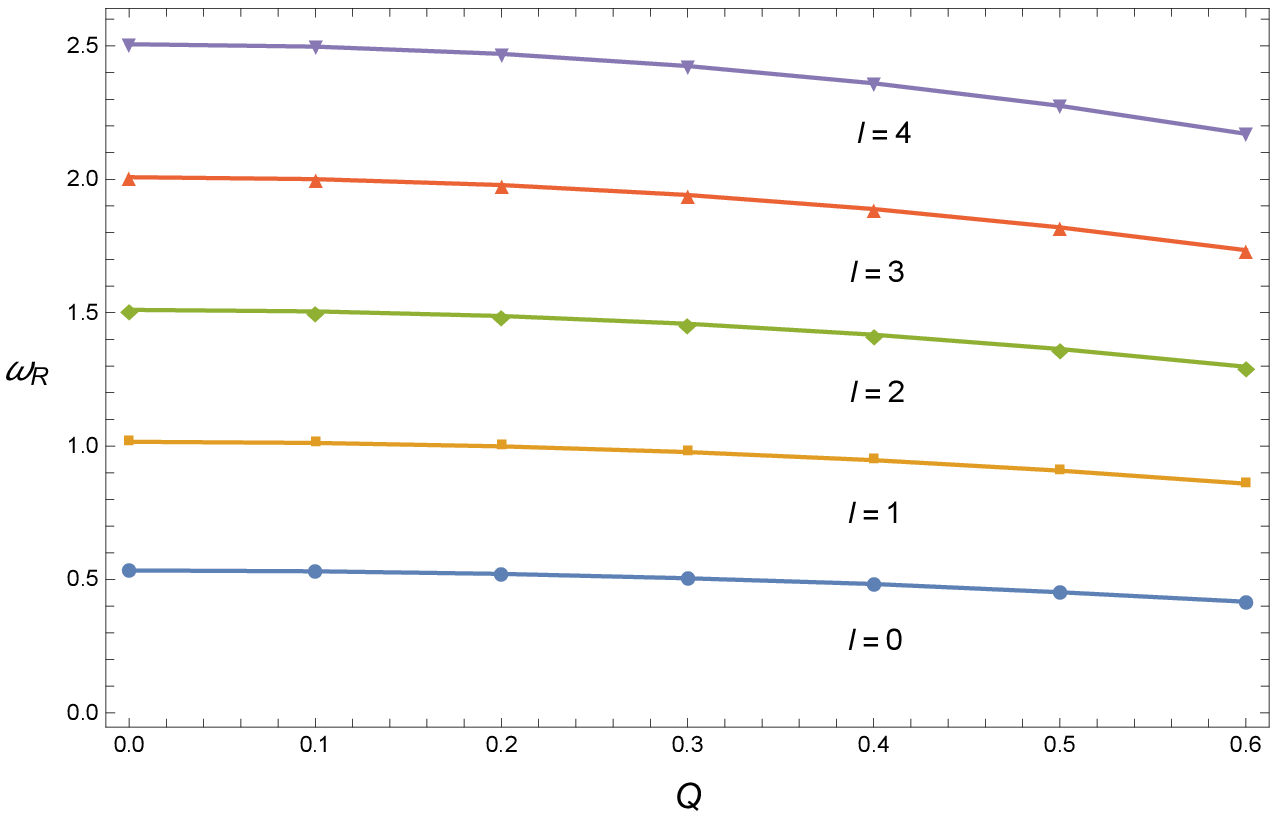}\label{fig:5dEYMRe}  \quad \includegraphics[width=0.45\linewidth]{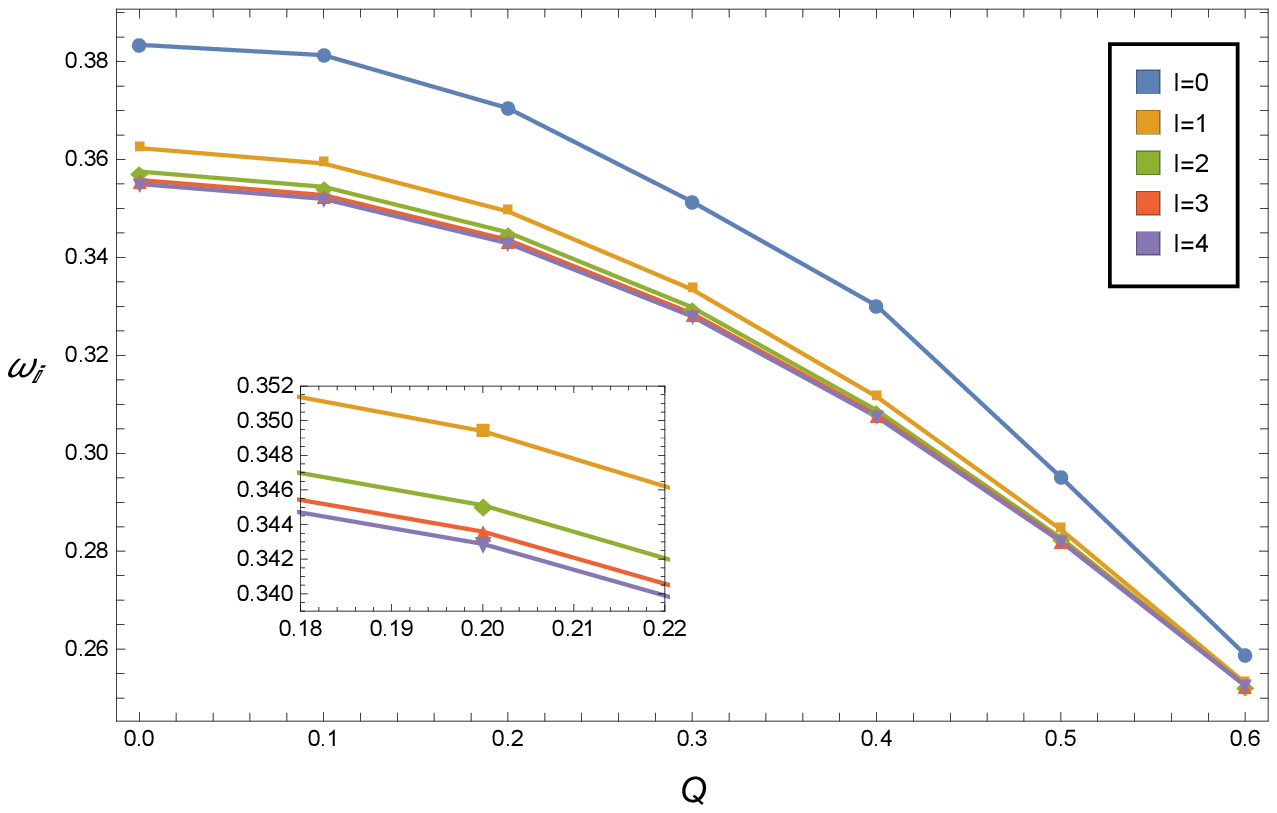} \label{fig:6dEYM(e}
	\caption{The real parts (left) and imaginary parts (right) of QNM frequencies as a function of the charge $Q$ at the angular quantum number $l$ from zero to four in the fundamental mode $(n=0)$ of the five-dimensional EYM black hole.
	} \label{fig:5DReIm}
\end{figure}

For the sake of intuition, the data in Table~\ref{tab:table1} are plotted in Fig.~\ref{fig:5DReIm}.
% Above we have only shown the modes with $l = 2$, and the results of different values of $l$ are listed in Table~\ref{tab:table1}.

We can see from Fig.~\ref{fig:5DReIm} that both the real and imaginary parts of the quasinormal mode frequencies at different angular quantum numbers have almost the same charge-dependent behavior, i.e., they decrease as the charge increases.  However, the significant difference between the real and imaginary parts of frequencies is that the former increases while the latter decreases when the angular quantum number $l$ increases, which shows~\cite{GTHH09} the same dependence on the angular quantum number as that in the Schwarzschild-AdS spacetime. The timescale $\tau$ for approaching to a thermal equilibrium can be characterized~\cite{GTHH09} by the inverse of imaginary parts of frequencies, $\tau=1/\omega_{\rm I}$. Thus we can see from the right diagram of Fig.~\ref{fig:5DReIm} that the imaginary parts of QNM frequencies show the similar dependence on the angular quantum number to that in the Schwarzschild-AdS spacetime in which the timescale increases for the modes with an increasing $l$.

%This phenomenon is understandable in terms of the AdS/CFT correspondence because the timescale for approaching to a thermal equilibrium increases for the modes with increasing $l$ and $Q$, see the right diagram of Fig.~\ref{fig:5DReIm}.

\subsection{The case of \boldmath$D>5$}\label{sec:d>5}
Now we pay attention to the dependence of QNM frequencies on the number of dimensions in EYM black holes, i.e., how the spacetime dimensions affect the quasinormal behavior of EYM black holes. By taking the same strategy as that for the case of $D=5$ in Sec.~\ref{sec:d=5}, we can obtain the expected QNM frequencies that are listed in Table~\ref{tab:table2}. Here we only focus on the the fundamental mode with $l=2$ for different values of charges $Q$ and spacetime dimensions $D$.

\begin{figure}[htbp]
	\centering
	\includegraphics[width=0.45\linewidth]{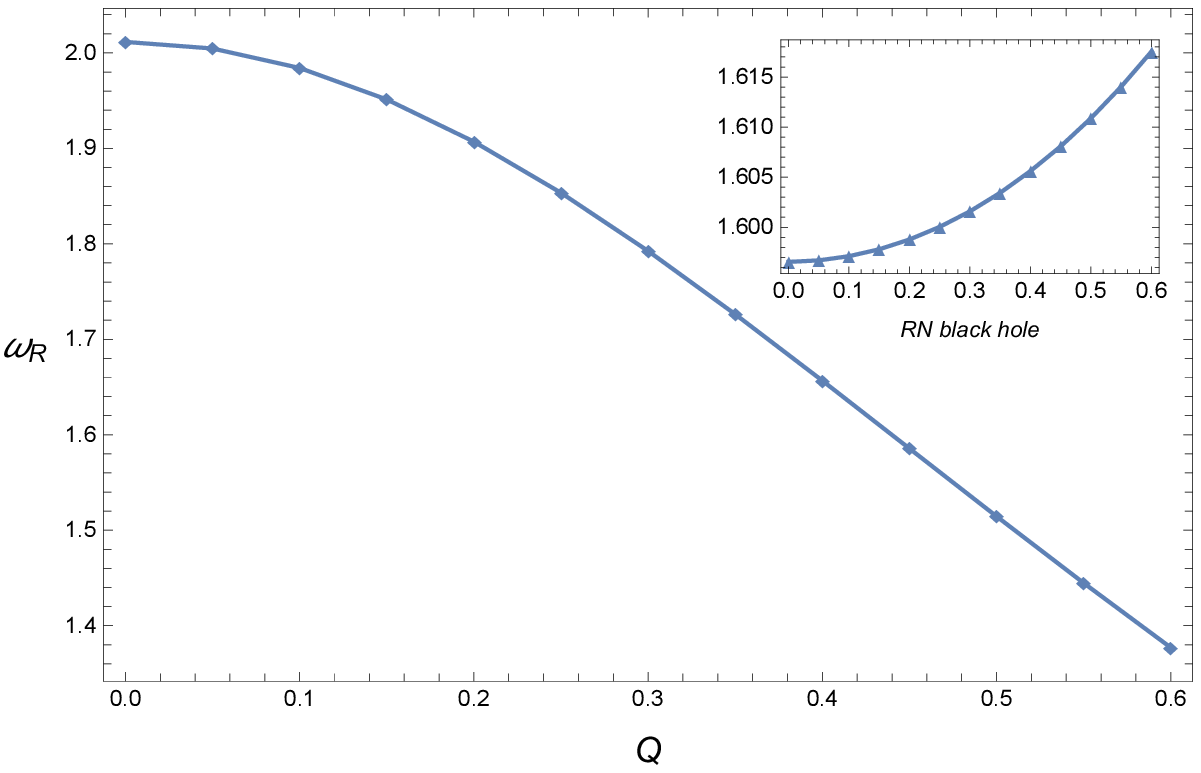}\label{fig:Re6d} \quad \includegraphics[width=0.45\linewidth]{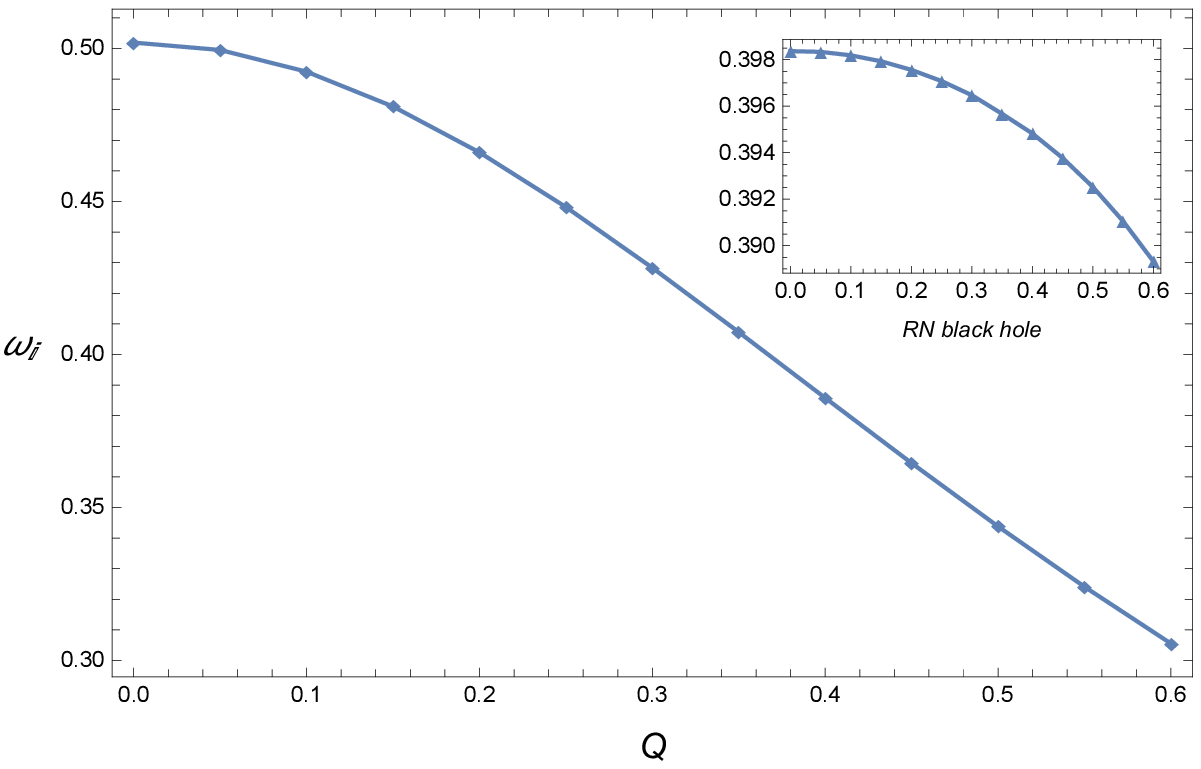} \label{fig:Im6d}
	
	\caption{The behaviors of real parts (left) and imaginary parts (right) of QNM frequencies in the fundamental mode $(n=0, l=2)$ with respect to the charge $Q$ for the EYM black hole and RN black hole in six dimensions.
	} \label{fig:6dReIm}
\end{figure}

In order to analyze in detail the behaviors of QNMs with respect to the gauge charge $Q$ in the EYM black hole, we compute more data for the fundamental mode ($n=0$) with the angular quantum number $l=2$ in $D=6$ as an example besides those in Table~\ref{tab:table2}, and plot the frequencies with respect to the charges in Fig.~\ref{fig:6dReIm}, where the case of the RN black hole is attached for comparison. 
We can see from Fig.~\ref{fig:6dReIm} that the dependence of real parts on charge is contrary to that of the six-dimensional RN black hole but the dependence of imaginary parts on charge consists with that of the six-dimensional RN black hole. By comparing Fig.~\ref{fig:5DRe} with Fig.~\ref{fig:6dReIm}, we can clearly see that the dependence of QNMs on charge in the six-dimensional EYM black hole is same as that in the five-dimensional one. This shows that there is no change in dependence of quasinormal mode frequencies on charge from five to six dimensions, which further implies that the logarithmic term that appears in five dimensions does not change such a dependence.  

\begin{table}[H]
	\caption{\label{tab:table2}
		The fundamental QNM frequencies $(n=0, l=2)$ of a massless scalar field perturbation for different values of $Q$ and $D$ in EYM black holes.}
                 \vskip 2mm
                 \resizebox{\textwidth}{!}{
		\begin{tabular}{ c ccccc}
			\hline
			\hline
			\textrm{$Q$}&	\textrm{$D=6$}&$D=7$&$D=8$&$D=9$&$D=10$\\
			\hline
			0.5 &1.514521-0.343802$i$&2.029577-0.453178$i$&2.479345-0.541278$i$&2.904600-0.617453$i$&3.316789-0.685071$i$\\
			0.6 &1.377058-0.305522$i$&1.875172-0.402034$i$&2.303481-0.476524$i$&2.703905-0.538803$i$&3.088930-0.592314$i$\\
			0.7 &1.250634-0.272229$i$&1.722885-0.355579$i$&2.124028-0.416261$i$&2.495122-0.464286$i$&2.848827-0.503528$i$\\
			0.8 &1.137836-0.243946$i$&1.579234-0.315325$i$&1.949849-0.363594$i$&2.288979-0.399086$i$&2.609131-0.425402$i$\\
			0.9 &1.038855-0.220126$i$&1.447956-0.281418$i$&1.787354-0.319631$i$&2.094371-0.345121$i$&2.381355-0.362000$i$\\
			1.0 &0.952647-0.200059$i$&1.330450-0.253212$i$&1.640160-0.283910$i$&1.917231-0.302592$i$&2.174113-0.313548$i$\\
			1.1 &0.877694-0.183068$i$&1.226518-0.229771$i$&1.509348-0.255100$i$&1.760130-0.269454$i$&1.991046-0.277319$i$\\
			1.2 &0.812412-0.168574$i$&1.135086-0.210166$i$&1.394266-0.231706$i$&1.622554-0.243457$i$&1.831985-0.249863$i$\\
			\hline
			\hline
		\end{tabular}}

\end{table}

\begin{figure}[H]
	\centering
	\includegraphics[width=0.45\linewidth]{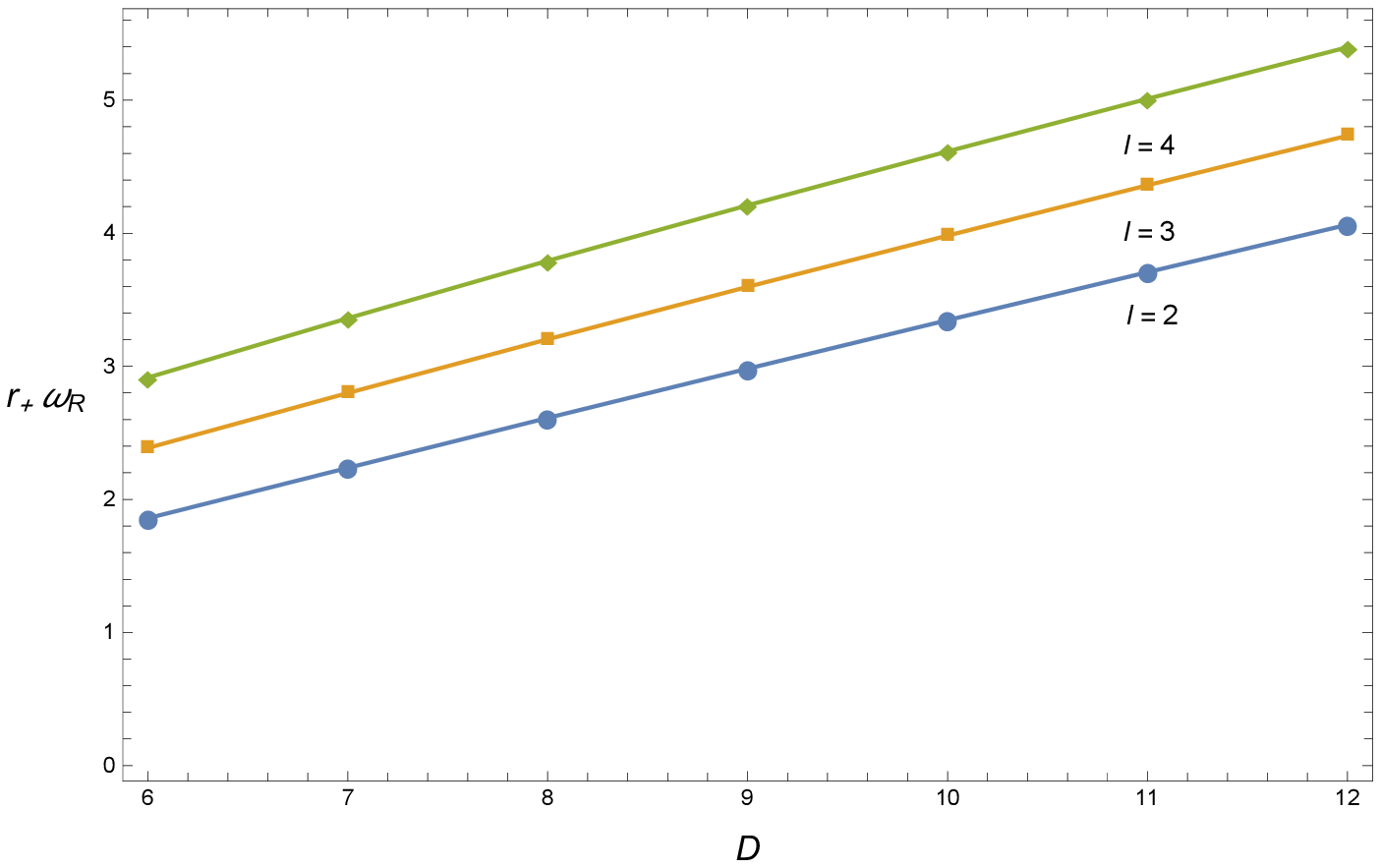}\label{fig:Re-d} \quad \includegraphics[width=0.458\linewidth]{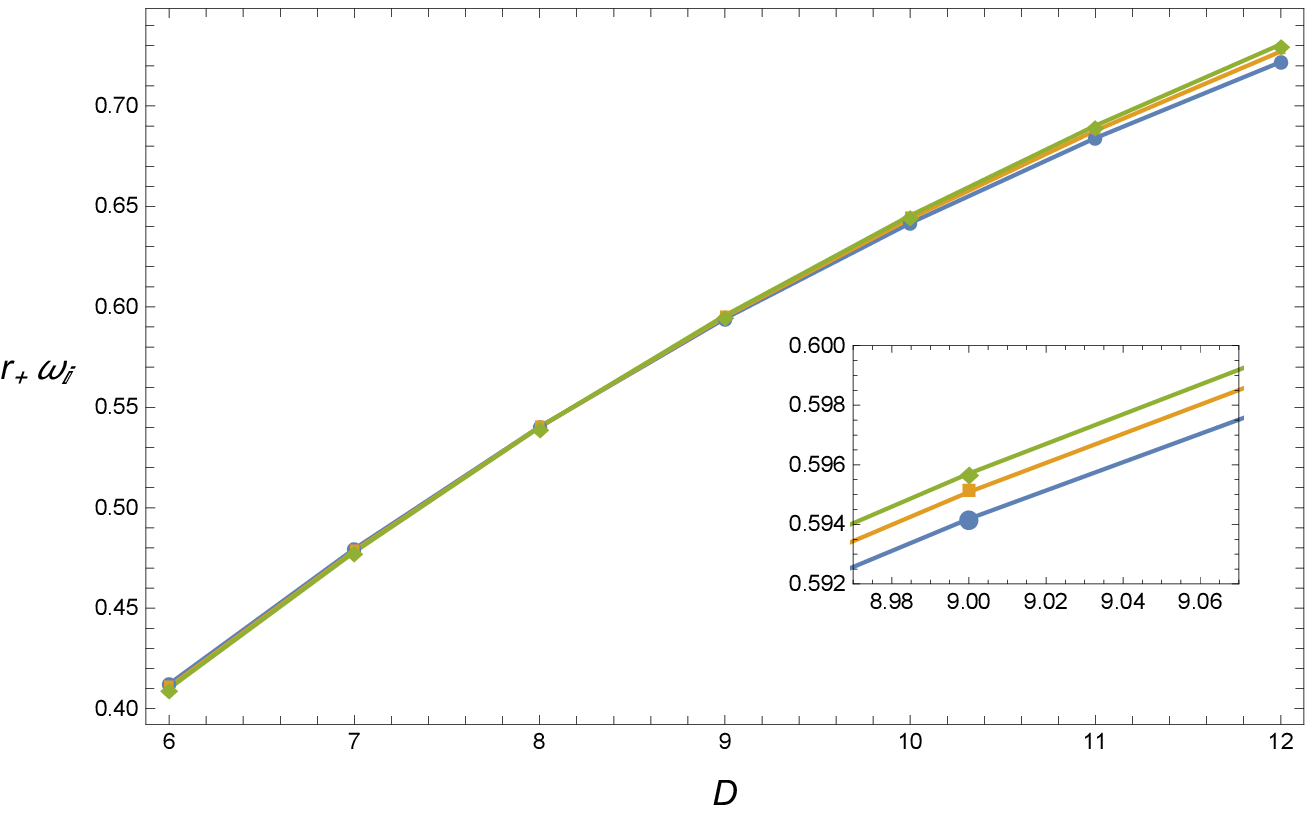} \label{fig:Im-d}
	
	\caption{Dependence of $r_{+}\omega_{\rm R}$ (left) and $r_{+}\omega_{\rm I}$ (right) on the number of dimensions $D$ for the angular quantum number $l$ from two to four in the fundamental mode ($n=0$) with a fixed charge $Q=0.6$.
	} \label{fig:6d-12dRe-d}
\end{figure}

In order to investigate the dependence of the QNMs on the number of dimensions $D$ in EYM black holes,
we plot Fig.~\ref{fig:6d-12dRe-d} showing the product of quasinormal frequencies and horizon radii with respect to the number of dimensions for the values of angular quantum numbers $l=2,3,4$, where we have taken the data on the second row in Table~\ref{tab:table2} and supplied the other data for $D=11$ and $D=12$.

%we plot Fig.~\ref{fig:6d-12dRe-d} showing the real and imaginary parts of frequencies with respect to dimensions for the values of angular quantum numbers $l=2,3,4$, where we have taken the data on the second row in Table~\ref{tab:table2} and supplied the other data for $D=11$ and $D=12$. 

We can see from Fig.~\ref{fig:6d-12dRe-d} that both $r_{+}\omega_{\rm R}$ and $r_{+}\omega_{\rm I}$ increase when $l$ increases, where the variations of the $r_{+}\omega_{\rm I}$ are very small for different values of $l$.
%differ from the five-dimensional case  and these profile almost overlap for different values of $l$. 
It is known~\cite{RAK08} that the real part of QNM frequencies is approximately proportional to the ratio of the spacetime dimension to the horizon radius, $D/r_{+}$, in high dimensional Schwarzschild black holes. 
Here we conclude from the left diagram of Fig.~\ref{fig:6d-12dRe-d} that the relation between $r_{+}\omega_{\rm R}$ and $D$ is almost  linear, which is similar to the situation in high dimensional Schwarzschild black holes. As a result, we write such a relation after giving the coefficients,
\begin{eqnarray}
\omega_{\rm R}\sim 0.328D/r_{+}\qquad (l=2),\\
%\end{eqnarray}
%\begin{eqnarray}
\omega_{\rm R}\sim 0.398D/r_{+}\qquad (l=3),\\
%\end{eqnarray}
%\begin{eqnarray}
\omega_{\rm R}\sim 0.468D/r_{+}\qquad (l=4).
\end{eqnarray}

\begin{figure}[H]
	\centering
	\includegraphics[width=0.45\linewidth]{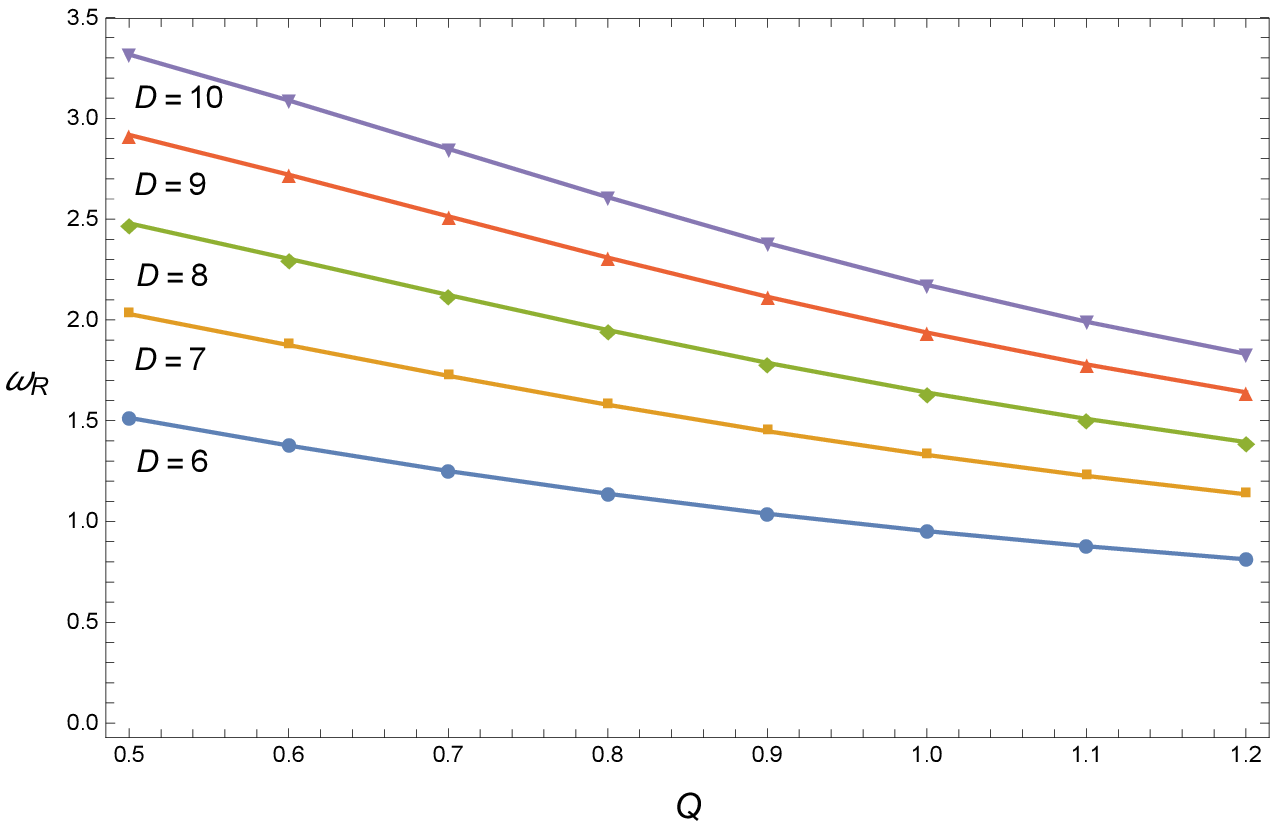} \label{fig:Re6d-10d} \quad \includegraphics[width=0.45\linewidth]{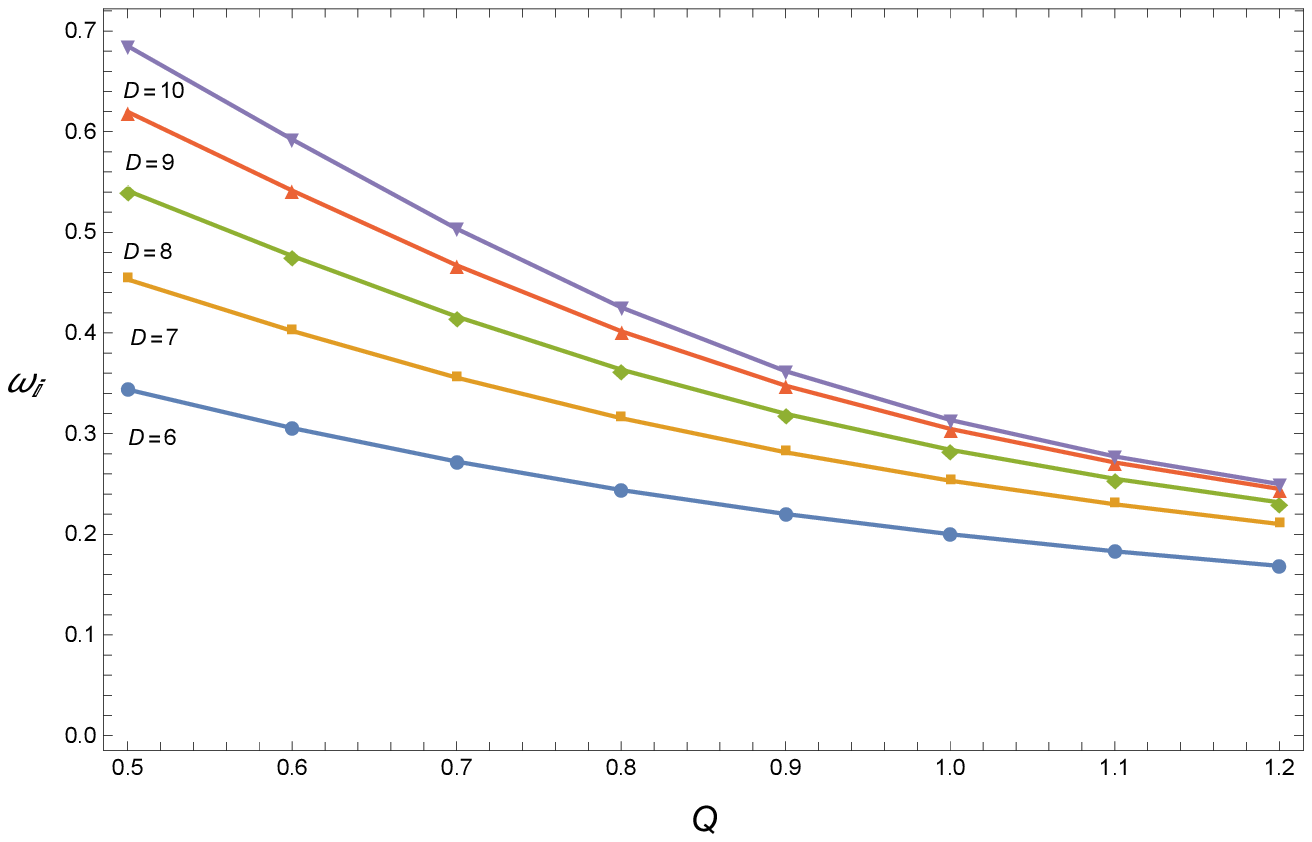} \label{fig:Im6d-10d}
	
	\caption{The real parts (left) and imaginary parts (right) of QNM frequencies in the fundamental mode $(n=0, l=2)$ as a function of the charge $Q$ for different values of $D$.
	} \label{fig:6d-10dReIm}
\end{figure}

For the sake of intuition, the data in Table~\ref{tab:table2} are plotted in Fig.~\ref{fig:6d-10dReIm}.

We can see from Fig.~\ref{fig:6d-10dReIm} that both the real and the imaginary parts of quasinormal mode frequencies decrease when the charge increases in the case of $D>5$, which is similar to the charge-dependent behavior in the case of $D=5$, see also Fig.~\ref{fig:5DReIm}.
% i.e.~it is progressively less for larger values of charge in different dimensions. 
Moreover, an increasing dimension with a fixed $Q$ leads to an interesting effect that both the real and imaginary parts are increasing.\footnote{For an increasing angular quantum number in the five-dimensional EYM black hole, the corresponding effect is different. That is, the real parts are ascending while the  imaginary parts are descending, see Fig.~\ref{fig:5DReIm} for the details.} In more detail, it is quite obvious from the right diagram of Fig.~\ref{fig:6d-10dReIm} that the increments of the imaginary parts with respect to the number of  dimensions decrease when the charge is increasing. In conclusion, the EYM black holes oscillate at larger frequencies but in shorter periods of damping time in higher spacetime dimensions.

\subsection{ Varying charge and numerical convergence}\label{sec:limit}

\begin{figure}[htbp]
	\centering
	\includegraphics[width=0.65\linewidth]{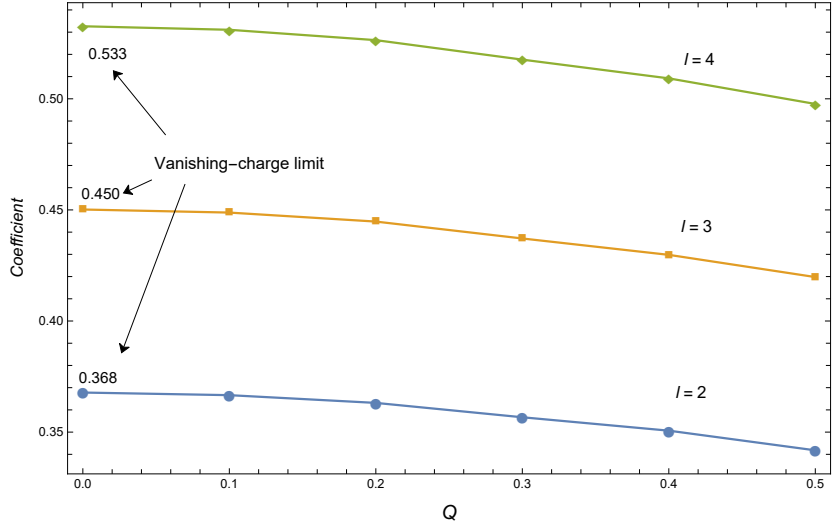}   
	\caption{The behaviors of the coefficients with respect to the charge for various values of angular quantum numbers.
	} \label{fig:CQ}
\end{figure}

In the above subsection we obtain the relation between $\omega_{\rm R}$ and $D/r_{+}$ for the fixed charge $Q=0.6$. In this subsection, we investigate the behavior of the coefficients of the relation for a varying charge from $0.5$ to zero and the numerical convergence of thirteen WKB orders as well. It is worth noting that  the mass term is an integration constant in the high dimensional EYM black holes, but in the high dimensional Schwarzschild black holes it was parameterized~\cite{RAK08,VCLY03}  to be dimension dependent. This makes the QNM frequencies of EYM black holes different from that of Schwarzschild black holes when the charge of EYM black holes approaches to zero. There are two choices~\cite{RAK08,VCLY03} for mass in high dimensional Schwarzschild black holes: $m=1$ and $m=2$. If we choose $m=1$ and $Q\to 0$ in high dimensional EYM black holes,  we shall lose the precision of the WKB method and the imaginary parts of QNM frequencies by using the Pad\'e approximation. Such modes are no longer the quasinormal modes we are searching for. Alternatively, we choose $m=2$ and plot the behavior of the coefficients of the relation between $\omega_{\rm R}$ and $D/r_{+}$ when the charge changes from 0.5 to zero in Fig.~\ref{fig:CQ}. We can see from this figure that the coefficients increase with the decreasing of charge and reach their maxima when the charge is vanishing. As a result, we can read off the coefficients from the figure for a varying charge in the range of $0\le Q \le 0.5$.

\begin{figure}[htbp]
	\centering
	\includegraphics[width=0.45\linewidth]{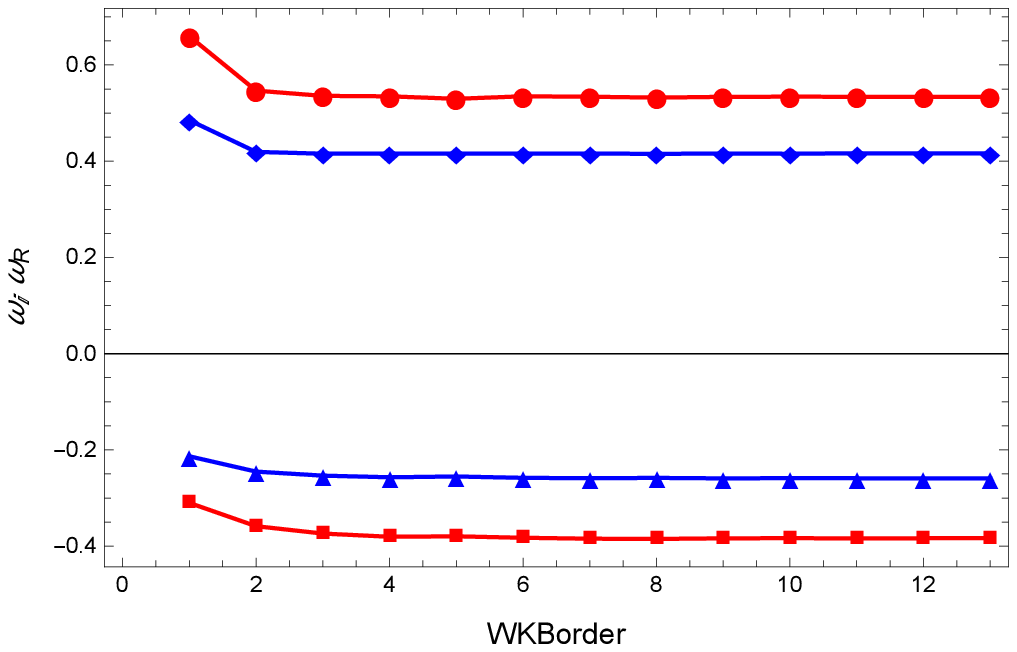}    \quad \includegraphics[width=0.45\linewidth]{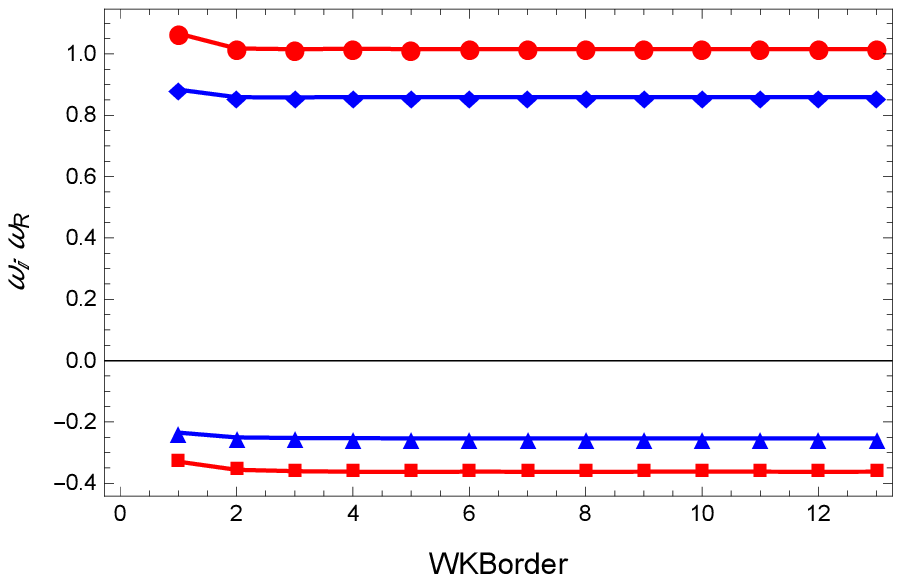}
	\caption{The convergence of $\omega_{\rm R}$ (top) and $\omega_{\rm I}$ (bottom) with respect to the WKB order number for the fundamental modes with $l=0$ (left diagram) and $l=1$ (right diagram) when $Q=0$ (red) and $Q=0.6$ (blue).
	}\label{fig:RIO}
\end{figure}

\begin{figure}[htbp]
	\centering
	\includegraphics[width=0.45\linewidth]{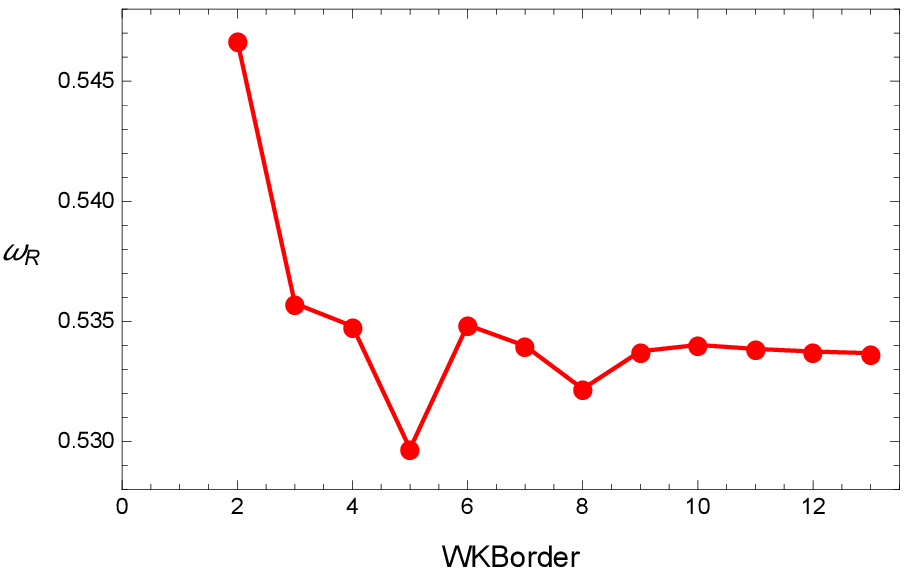} \quad \includegraphics[width=0.45\linewidth]{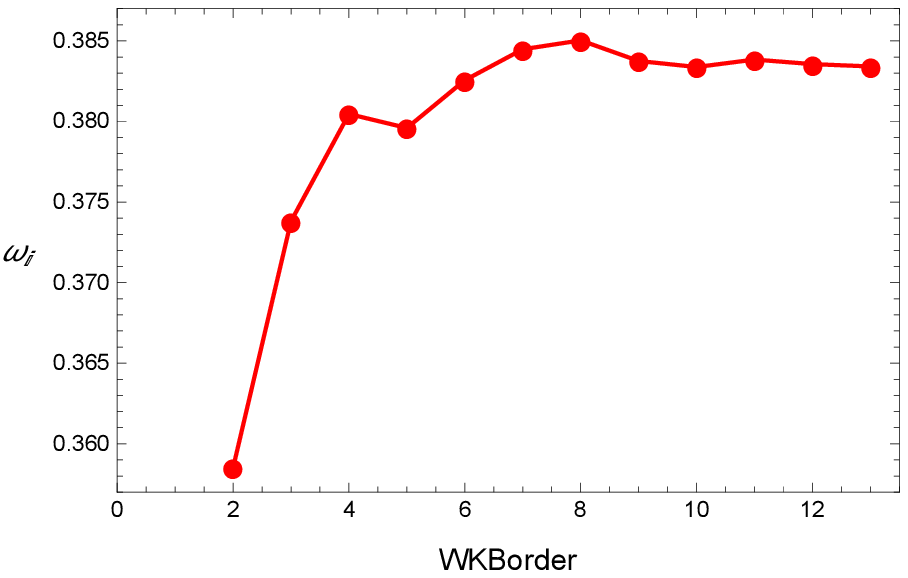}
	\includegraphics[width=0.45\linewidth]{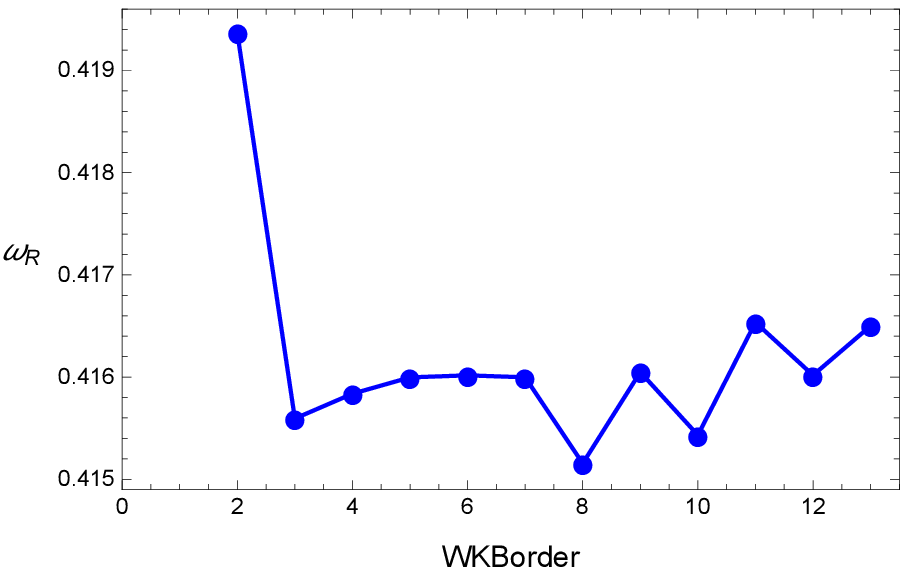} \quad \includegraphics[width=0.45\linewidth]{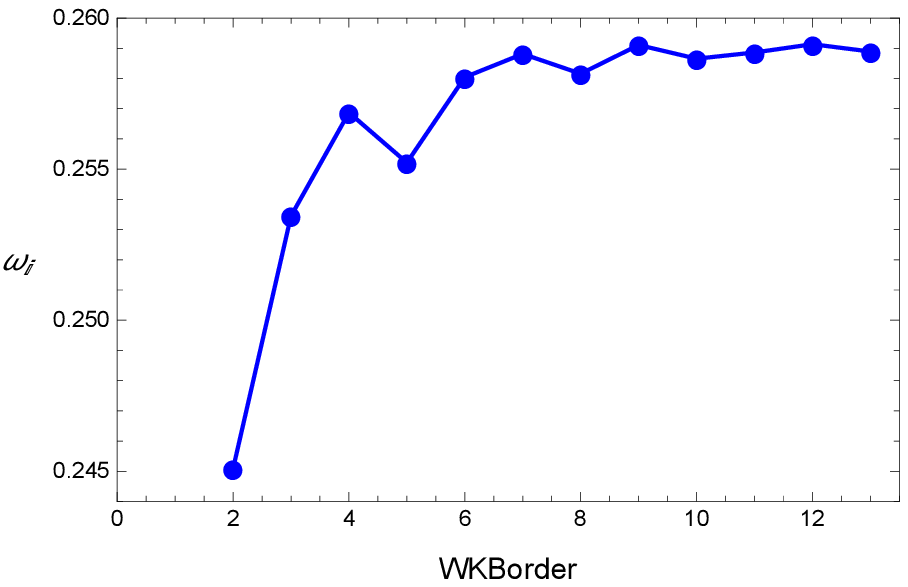}
	\caption{The convergence of $\omega_{\rm R}$ and $\omega_{\rm I}$ with respect to the WKB order number for the fundamental modes with $l=0$ at $Q=0$ (red) and $Q=0.6$ (blue).
	} \label{RIO0}
\end{figure}

\begin{figure}[htbp]
	\centering
	\includegraphics[width=0.45\linewidth]{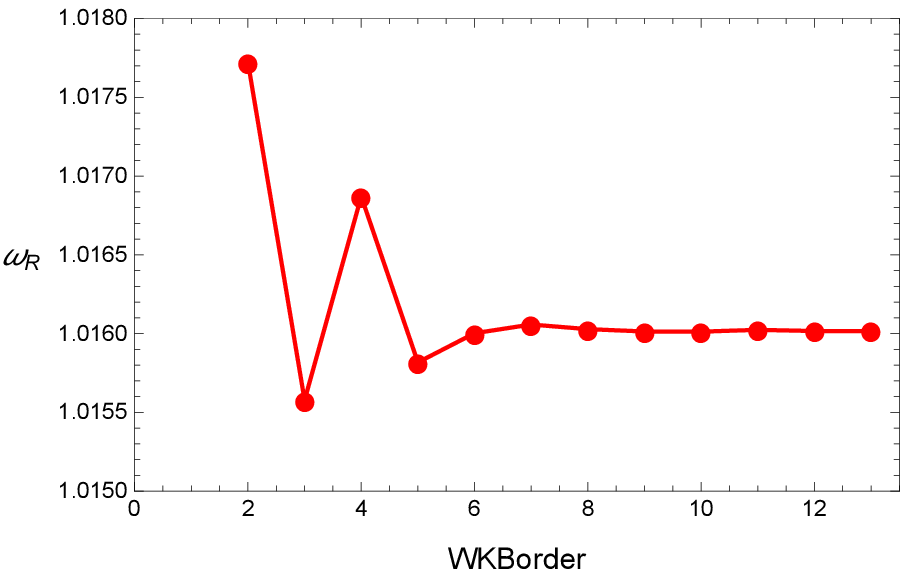} \quad \includegraphics[width=0.45\linewidth]{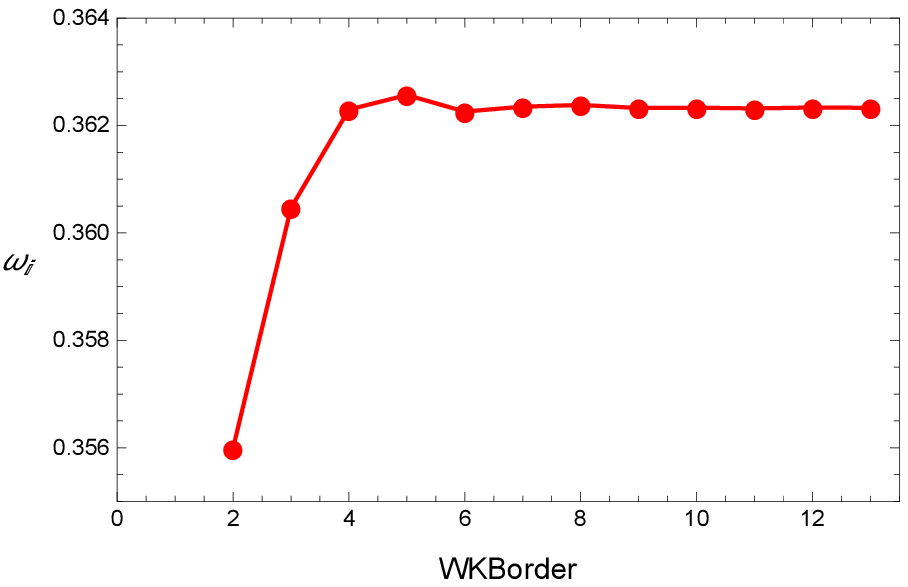}
	\includegraphics[width=0.45\linewidth]{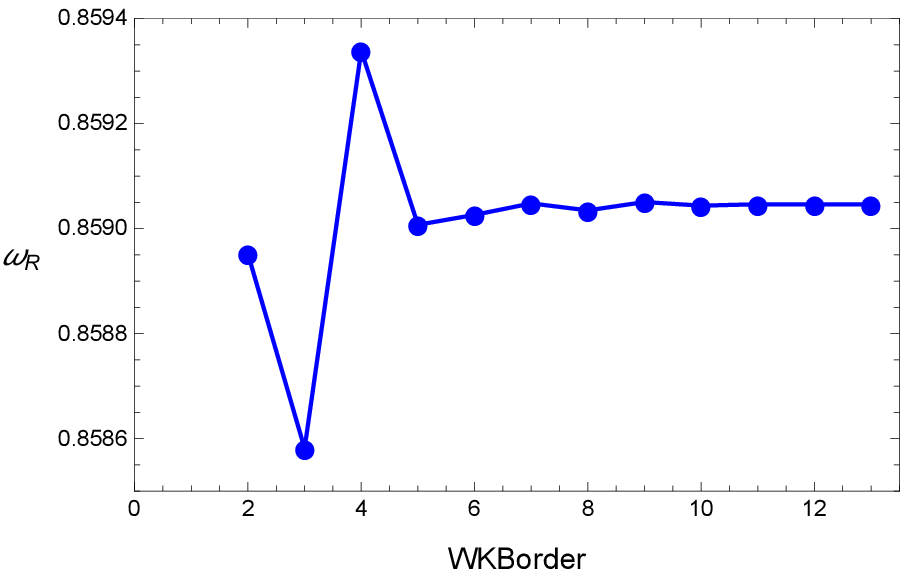} \quad \includegraphics[width=0.45\linewidth]{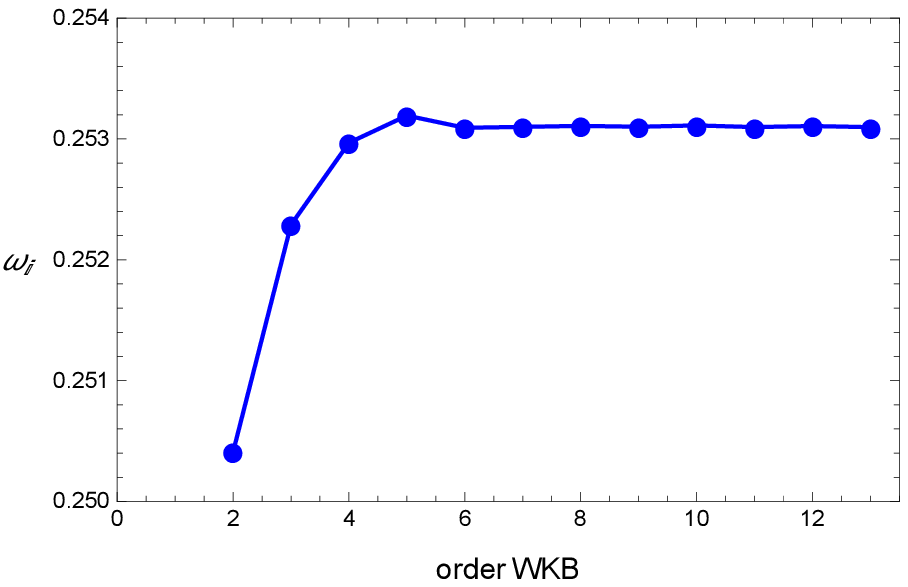}
	\caption{$\omega_R$ and $\omega_I$ The convergence of with respect to the WKB order number for the fundamental modes with $l=1$ at $Q=0$ (red) and $Q=0.6$ (blue).
	} \label{RIO1}
\end{figure}

We show the numerical convergence with respect to the WKB order number from one to thirteen in Figures~\ref{fig:RIO}, \ref{RIO0} and \ref{RIO1}. 

In Fig.~\ref{fig:RIO}, we plot the convergence of QNM frequencies with respect to the number of WKB orders in the five-dimensional spacetime for the fundamental modes with $l=0$ (see the left diagram of Fig.~\ref{fig:RIO}) and $l=1$ (see the right diagram of Fig.~\ref{fig:RIO}). We can see that the greatest deviation occurs between the first and second WKB orders.

To show the numerical convergence more intuitively, we ignore the first order and then plot the numerical convergence with respect to the number of WKB orders from two to thirteen in Fig.~\ref{RIO0} for the fundamental modes with $l=0$ and in Fig.~\ref{RIO1} for the fundamental modes with $l=1$. 

We can clearly see from Fig.~\ref{RIO0} that the fundamental modes with $l=0$ show a bad convergence and have a hard time to converge to the required accuracy even though we have used the improved WKB method up to the 13th order.
In Ref.~\cite{RAK08}, the sixth order WKB method was applied to the fundamental modes with $l=0$ and $l=1$, and a bad accuracy and a considerable relative error were found, where the relative error was larger for a higher dimension. From Fig.~\ref{RIO1} we can see that a significant improvement of convergence starts at order six for the fundamental modes with $l=1$. This shows that the 13th order WKB method has a better numerical convergence than that of the sixth order WKB method. In spite of the feature, we find that the QNM frequencies of the fundamental modes with $l=0$ and $l=1$ still fail to converge to the required accuracy in high dimensional spacetimes. Therefore, we only focus on the fundamental modes with $l\ge2$ throughout this paper. In summary, the 13th order WKB method is more suitable for studying the high dimensional black hole background spacetimes because it has a better numerical convergence than that of the 6th order WKB method.

\section{ Conclusions}\label{sec:conclusions}

By using the improved WKB method, we have computed the scalar QNMs of Einstein-Yang-Mills black holes in different spacetime dimensions. 
The logarithmic term in the metric function of the five-dimensional Einstein-Yang-Mills black hole prevents the Hawking temperature from diverging when the horizon radius shrinks, but it keeps the charge-dependent behavior of the QNMs unchanged. In any number of dimensions larger than four, the real and imaginary parts of quasinormal mode frequencies decrease with the increasing of charge, which is different from the situation in the real parts of RN black holes. For the fundamental modes of the EYM black holes, the real parts of QNM frequencies are approximately proportional to the ratio of the number of dimensions to the radius of horizons, $D/r_{+}$, in the range of $0\le Q\le 0.6$. However, the imaginary parts are almost same for different values of angular quantum numbers, i.e. such tiny variations of the imaginary parts imply that the Einstein-Yang-Mills black holes approach to a thermal equilibrium at a nearly same damping timescale for different angular quantum numbers. 
As for the tendency related to the number of dimensions, our results show that the EYM black holes oscillate at larger frequencies but decay in shorter timescales in higher dimensional spacetimes. A pre-existing open question~\cite{RAK08} in the research of QNMs is whether the relationship between QNMs and the number of dimensions will be present or not for a general background spacetime beyond high dimensional Schwarzschild spacetimes. We confirm that the relationship is well-behaved in the EYM spacetimes with the $SO(D-1)$ gauge symmetry. 
%In addition, our work can provide a favorable evidence that QNM frequencies are determined only by parameters of black holes such as charge, mass and spin. 
Since the number of spacetime dimensions is related to the gauge group such as $SO(D-1)$ here, whether such a relationship depends on a non-abelian gauge field and its related gauge symmetry remains to be further investigated.

\section*{Acknowledgments}

The authors would like to thank the anonymous referee for the helpful comments that improve this work greatly. This work was supported in part by the National Natural Science Foundation of China under grant No. 11675081.

\end{document}